\documentclass[referee]{sn-jnl} 
\usepackage{amsmath}
\usepackage{amssymb}
\usepackage{geometry}
\usepackage{graphicx}
\usepackage{comment,xurl,hyperref,multirow,amsfonts,appendix}
\DeclareUnicodeCharacter{0308}{\"}

\title[Pattern formation in AD]{Pattern formation in a Reaction-Diffusion Model for Amyloid-\texorpdfstring{$\beta$}{beta} and Tau Interactions in Alzheimer's Disease}

\author*[1]{\fnm{Sun} \sur{Lee}}\email{skl5876@psu.edu}
\author[1]{\fnm{Wenrui} \sur{Hao}}\email{wxh64@psu.edu}

\affil*[1]{\orgdiv{Department of Mathematics}, \orgname{The Pennsylvania State University}, \orgaddress{\street{201 Old Main}, \city{University Park}, \postcode{16802}, \state{PA}, \country{USA}}}

\begin{document}

\abstract{Alzheimer's disease (AD) is characterized by the accumulation of Amyloid-$\beta$ ($A\beta$) plaques and hyperphosphorylated Tau proteins. However, many individuals exhibit substantial $A\beta$ and Tau pathology without developing dementia, suggesting that disease progression may depend not only on pathological burden but also on the spatial organization of these proteins. Motivated by this observation, we adapt  Gray-Scott reaction-diffusion model to investigate pattern formation arising from the interactions between $A\beta$ and Tau.
To systematically identify stable spatial configurations, we employ a Companion-Based Multi-Level Finite Element Method (CBMFEM) on both two-dimensional domains and anatomically realistic cortical surface meshes. Numerical simulations reveal a rich landscape of multiple steady-state solutions, which are subsequently classified into representative pattern phenotypes using principal component analysis and clustering techniques. The results demonstrate that the coupled $A\beta$--Tau system admits numerous stable spatial patterns rather than a single pathological endpoint.
These findings provide a potential mathematical framework for understanding the heterogeneity of Alzheimer's disease and the existence of cognitively resilient individuals despite significant pathological burden. More broadly, the proposed framework suggests a pattern-based therapeutic paradigm in which disease dynamics are guided toward favorable stable states rather than solely targeting the elimination of pathological proteins.
}
\maketitle

\section{Introduction}

Alzheimer's disease (AD) is the most common cause of dementia worldwide and is characterized by progressive cognitive decline, memory impairment, and neuronal loss. The two principal neuropathological hallmarks of AD are the extracellular deposition of Amyloid-$\beta$ ($A\beta$) plaques and the intracellular accumulation of hyperphosphorylated Tau proteins in the form of neurofibrillary tangles \cite{braak1991neuropathological,hardy1992alzheimer,selkoe1991amyloid}. The amyloid cascade hypothesis proposes that abnormal accumulation of $A\beta$ initiates a series of pathological events that ultimately promote Tau pathology, synaptic dysfunction, and neurodegeneration \cite{hardy2002amyloid,hardy1992alzheimer}. Consequently, substantial efforts have been devoted to developing anti-$A\beta$ and anti-Tau therapies aimed at reducing pathological protein burden and slowing disease progression.

Despite the central role of $A\beta$ and Tau in AD pathology, a longstanding paradox remains unresolved. Numerous neuropathological and imaging studies have identified individuals who harbor substantial amyloid plaques and Tau pathology yet remain cognitively intact throughout life \cite{price2009neuropathology,sperling2011toward}. Conversely, some individuals with relatively modest pathological burden exhibit significant cognitive impairment. These observations suggest that the absolute quantity of pathological proteins alone may be insufficient to determine clinical outcome. Instead, additional mechanisms governing the interaction and organization of $A\beta$ and Tau likely influence whether pathology progresses toward dementia or remains relatively benign.

One possible explanation is that the coupled dynamics of $A\beta$ and Tau admit multiple stable states. Rather than converging to a single pathological endpoint, the disease process may evolve toward distinct spatial configurations characterized by different levels of stability and neurotoxicity. Some configurations may promote widespread propagation of pathology and neuronal damage, while others may correspond to relatively stable states that limit further disease progression \cite{hao2022optimal,lee2025optimal}. Such a perspective is consistent with the growing recognition that resilience to Alzheimer's disease is influenced not only by pathological burden but also by the spatial distribution and interaction of pathological processes across brain networks \cite{jack2018nia,stern2002cognitive}.

The existence of multiple stable pathological configurations has important clinical implications. Current therapeutic strategies largely focus on eliminating $A\beta$ and Tau from the brain. However, if certain stable patterns are compatible with preserved cognitive function, complete removal of pathology may not always be necessary. Instead, treatment decisions should be informed by whether a patient's disease trajectory is approaching an aggressive pathological state or a relatively benign stable configuration. Such an approach could help identify patients who would benefit most from intervention while reducing unnecessary treatment in individuals whose pathology remains stable.

Furthermore, the existence of favorable stable states suggests an alternative therapeutic paradigm. Rather than seeking complete eradication of pathological proteins, interventions could be designed to redirect disease dynamics toward stable and less harmful configurations. In this setting, the therapeutic objective becomes the identification and stabilization of desirable disease states. This perspective transforms Alzheimer's disease from a purely elimination-based problem into a control problem, where the goal is to steer pathological dynamics toward favorable attractors in the underlying biological system.

Mathematical modeling provides a natural framework for investigating these questions \cite{petrella2019computational,petrella2024personalized,rabiei2025data,wang2026learning,zheng2022data}. Reaction-diffusion systems have been widely used to study pattern formation in biological, chemical, and ecological systems, where local interactions coupled with diffusion generate multiple steady states, self-organized structures, and Turing-type patterns \cite{cross1993pattern,murray2003mathematical,turing1952chemical}. Similar approaches have been successfully applied to describe protein propagation and disease progression in neurodegenerative disorders \cite{raj2012network,weickenmeier2019brain}. Within this framework, the interactions between $A\beta$ and Tau can be viewed as a nonlinear pattern-forming process whose stable spatial solutions may correspond to distinct disease phenotypes \cite{bertsch2017alzheimer,fornari2019prion,hao2016mathematical,pearson1993complex,weickenmeier2019physics}.

Motivated by these observations, we adapted Gray-Scott model describing the coupled dynamics of $A\beta$ and Tau and investigate its pattern formation behavior on anatomically realistic brain domains \cite{hao2025pattern,hao2020spatial}. Rather than focusing solely on disease propagation, our objective is to systematically characterize the spectrum of stable spatial patterns supported by the system. We hypothesize that some of these patterns may represent biologically plausible stable states associated with limited disease progression, thereby providing a potential explanation for the existence of cognitively normal individuals with substantial AD pathology. Ultimately, this work aims to establish a mathematical foundation for pattern-based patient stratification and to explore a new treatment paradigm in which therapeutic interventions guide disease dynamics toward favorable stable states.

The remainder of this paper is organized as follows. In Section 2, we develop a reaction-diffusion model describing the interactions between Amyloid-$\beta$ and Tau proteins and discuss its biological interpretation in the context of Alzheimer's disease progression. Section 3 presents the estimation and normalization of diffusion parameters based on experimental measurements and physiological considerations. In Section 4, we introduce the computational brain domain and the Companion-Based Multi-Level Finite Element Method (CBMFEM) used to compute multiple steady-state solutions on anatomically realistic cortical surfaces. Section 5 presents numerical simulations on both two-dimensional benchmark domains and three-dimensional cortical manifolds, where we systematically identify and classify distinct pattern formation phenotypes. Finally, Section 6 summarizes the main findings and discusses their implications for understanding Alzheimer's disease resilience, patient stratification, and the development of future pattern-guided therapeutic strategies.

\section{Mathematical Modeling of A\texorpdfstring{$\beta$}{beta} and Tau Interactions}
A major pathological feature strongly associated with the progression of Alzheimer's disease (AD) is the accumulation and potential synergistic interaction of two neuropathological hallmarks: extracellular Amyloid-$\beta$ ($A\beta$) plaques and intracellular neurofibrillary tangles composed of hyperphosphorylated Tau proteins \cite{querfurth2010alzheimer}. While the exact etiology of AD remains a subject of ongoing scientific debate, one of predominant framework in the field is the amyloid cascade hypothesis \cite{hardy1992alzheimer, selkoe2016amyloid}. This hypothesis posits that $A\beta$ pathology occurs upstream, creating a neurotoxic environment that is suspected to facilitate and accelerate the misfolding and neocortical spread of Tau \cite{busche2020synergy,scheltens2021alzheimer}.

Taking this prominent biological premise as our starting point, we adopt a Gray-Scott type reaction-diffusion system to mathematically capture this inter-protein dynamic \cite{bertsch2017alzheimer,fornari2019prion,hao2016mathematical,pearson1993complex,weickenmeier2019physics}. Under the assumption of the amyloid cascade hypothesis, the Gray-Scott model is highly suitable because its core autocatalytic mechanism mathematically reflects the hypothesized $A\beta$-mediated proliferation of Tau \cite{jucker2013self}. In this framework, $A\beta$ is modeled as a requisite resource or catalyst that is consumed to drive the aggregation and propagation of Tau.
More specifically, let $u(\mathbf{x},t)$ represent the concentration of $A\beta$ and $v(\mathbf{x},t)$ represent the concentration of Tau, normalized such that their maximum theoretical concentrations correspond to $1$. Following the aforementioned hypothesis, the system dictates that $v$ will proliferate through an autocatalytic interaction with $u$, while $u$ will deplete as a result of this catalytic reaction. The spatiotemporal dynamics on the brain domain are governed by the following system of partial differential equations:
\begin{equation}\label{eq:1}
    \frac{\partial u}{\partial t} = D_u \Delta u + F(1 - u) - d u v^2
\end{equation}
\begin{equation}\label{eq:2}
    \frac{\partial v}{\partial t} = D_v \Delta v + d u v^2 - (F + k) v
\end{equation}

Each term in this Gray-Scott framework maps to a specific biological mechanism within the AD progression:

\begin{itemize}
    \item \textbf{Homeostasis and Production ($F(1 - u)$):} This term represents the natural production and steady-state maintenance of A$\beta$ in the brain, reflecting in vivo synthesis and baseline clearance dynamics \cite{bateman2006human}. 
    \item \textbf{Prion-like Autocatalysis ($d u v^2$):} The nonlinear interaction term represents the A$\beta$-facilitated misfolding of Tau. The quadratic dependence on $v$ ($v^2$) mathematically captures the templated aggregation process, where existing hyperphosphorylated Tau recruits and converts healthy proteins, scaled by the coupling coefficient $d$. This formulation is consistent with empirical findings that A$\beta$ enhances Tau-seeded pathologies \cite{he2018amyloid}.
    \item \textbf{Clearance and Degradation ($-(F + k)v$):} This linear degradation term accounts for the physiological clearance of Tau aggregates through neuroimmune responses \cite{bolos2017absence, yoo2024complement} and the glymphatic system \cite{harrison2020impaired}, where $k$ denotes the specific decay rate of misfolded Tau.
    \item \textbf{Differential Diffusion ($D_u > D_v$):} The diffusion coefficients, $D_u$ and $D_v$, describe the spatial spread of the proteins \cite{murray2003mathematical}. The constraint $D_u > D_v$ is a mathematical requisite for generating Turing-type spatial patterns \cite{turing1952chemical} within the Gray-Scott system \cite{pearson1993complex}. As derived in the subsequent section, normalized diffusion coefficients estimated from experimental literature yield $D_{u,norm} \approx 0.09$ for A$\beta$ \cite{danielsson2002translational} and $D_{v,norm} \approx 0.006$ for Tau \cite{janning2014single}. While these specific values are derived from localized experimental settings, the resulting order-of-magnitude difference aligns with the broader macroscopic observation: A$\beta$ diffuses through the extracellular space, whereas Tau pathology propagates more slowly via intracellular mechanisms and axonal pathways.
\end{itemize}

While numerous experimental studies have quantified the kinetic rates of A$\beta$ and Tau across various settings---ranging from in vitro assays and transgenic animal models to clinical cerebrospinal fluid (CSF) tracking \cite{bateman2006human, sato2018tau, yamada2015analysis}---directly mapping these scale-dependent, dimensional measurements to the abstract parameters of our dimensionless Gray-Scott system remains theoretically ambiguous. Consequently, rather than enforcing a direct numerical translation that may misrepresent the macroscopic system, the kinetic parameters in our model are determined phenomenologically to capture the qualitative dynamics of the disease. We set the production rate to $F = 0.4$, the decay parameter to $k = 0.65$, and the coupling coefficient to $d = 10$ in Eq. \eqref{eq:1}, \eqref{eq:2}. These dimensionless values were systematically chosen to ensure the reaction-diffusion system operates within the Turing instability regime. This mathematical calibration allows the model to transition from a homogenous steady state to stable, spatially heterogeneous patterns, which is characteristic of localized neurodegenerative pathology \cite{fornari2019prion}.

Having established the mathematical and biological framework of the Gray-Scott system for AD progression, the primary objective of this study is to systematically discover a comprehensive landscape of pattern formation solutions across the brain mesh. Rather than merely simulating a single transient progression, our goal is to compute the multiple steady-state solutions and Turing-type patterns that this nonlinear reaction-diffusion system can support.

Mathematically, uncovering a diverse array of solutions allows us to understand the underlying bifurcation structure and the ultimate spatial distribution of the proteins. Biologically, this carries significant implications. The varied pattern formation solutions are interpreted as theoretical representations of the potential end-states or ultimate pathological stages of the brain under different disease parameters. By identifying these multiple coexisting steady states, we aim to predict the potential long-term spatial distributions of A$\beta$ and Tau. 

Crucially, this comprehensive mapping reveals not only severe pathological stages but also more benign, stable configurations. In a clinical context where a patient's condition is actively deteriorating, identifying these stable alternatives becomes highly actionable. By applying optimal control theory, therapeutic interventions can be mathematically designed to steer the patient's biochemical dynamics away from a severe trajectory and guide the system toward a nearby, less aggressive steady state. While this approach may not completely reverse existing neurodegeneration, trapping the disease dynamics in a more benign state effectively stabilizes the condition or significantly slows further progression---a central paradigm in modern disease-modifying therapies for Alzheimer's disease. Therefore, capturing a broad spectrum of multiple solutions establishes the essential theoretical foundation for designing these targeted, optimal control-based treatments \cite{cummings2024alzheimer}.

\section{Diffusion Parameter estimations}
To simulate the progression of AD on a macroscopic brain mesh, the physical diffusion coefficients must be non-dimensionalized. We define a computational domain where the maximum characteristic length $L$ (the average anterior-posterior length of an adult human brain, approximately $16.7\text{ cm} = 0.167\text{ m}$) \cite{blinkov1968human} is scaled to $1$ unit. The characteristic time $T$ corresponds to $1$ year, which is defined as $365.25 \times 24 \times 3600 = 31,557,600\text{ s}$. The dimensionless diffusion coefficient $D_{norm}$ is calculated as:
\begin{equation}
    D_{norm} = \frac{D_{phys} \times T}{L^2}
\end{equation}

\subsubsection*{Assumption of Isotropic Surface Diffusion} 
In a strictly physiological context, diffusion within the human brain is highly heterogeneous and anisotropic, heavily directed by the structured white matter tracts. However, for the scope of this study, we restrict our computational domain to the two-dimensional cortical surface mesh representing the gray matter. Clinically, while AD is a highly complex disorder, the most prominent macroscopic burden of these specific proteinopathies manifests within the gray matter. While the precise spatiotemporal initiation differs—with $A\beta$ first appearing in the neocortex and Tau originating in the transentorhinal regions—both plaques and tangles predominantly accumulate and propagate within the cortical gray matter rather than the structured white matter tracts \cite{braak1991neuropathological, thal2002phases}. By restricting the domain to this surface manifold, we can reasonably approximate the diffusion process as isotropic and homogeneous. This critical simplification allows us to utilize uniform scalar diffusion coefficients ($D_u$ and $D_v$), providing a mathematically tractable baseline to rigorously isolate and study the fundamental pattern formation dynamics without the confounding variables introduced by spatial anisotropy.

\subsubsection*{Biological Corrections}
The raw experimental diffusion values, typically measured in aqueous solutions at room temperature, must be corrected to reflect the complex biological environment of the brain in vivo ($37^\circ\text{C}$ and tissue hindrance). According to Syková \& Nicholson (2008)~\cite{sykova2008diffusion}, diffusion in the brain's extracellular space is hindered by the geometric complexity of the tissue. The effective diffusion coefficient $D^*$ is related to the free diffusion coefficient $D$ by the tortuosity $\lambda$ ($D^* = D/\lambda^2$). Using the standard brain tortuosity $\lambda \approx 1.6$, the diffusion speed is reduced by a factor of $\lambda^2 \approx 2.56$. 

Furthermore, diffusion is temperature-dependent, following the Stokes-Einstein relation ($D \propto T/\eta$, where $\eta$ is dynamic viscosity). To correct experimental Amyloid-$\beta$ ($A\beta$) data measured at $25^\circ\text{C}$ to physiological body temperature ($37^\circ\text{C}$), we apply a temperature and viscosity correction factor $\alpha$:
\begin{equation}
    \alpha = \frac{T_{310.15K}}{T_{298.15K}} \times \frac{\eta_{H_2O, 25^\circ C}}{\eta_{H_2O, 37^\circ C}} \approx 1.04 \times 1.29 \approx 1.34
\end{equation}

{\bf Refined Normalization of Diffusion Coefficients}
Applying the aforementioned biological corrections, we calculate the effective diffusion coefficients for both $A\beta$ and Tau, followed by their non-dimensionalization for the computational model.

For Amyloid-$\beta$ ($u$), the baseline diffusion coefficient measured at $25^\circ\text{C}$ is $D_{u,raw} = 1.52 \times 10^{-10}\text{ m}^2/\text{s}$ \cite{danielsson2002translational}. To account for physiological body temperature and tissue hindrance, we apply both the temperature-viscosity correction ($\alpha \approx 1.34$) and the tortuosity penalty ($\lambda^2 \approx 2.56$) \cite{sykova2008diffusion}. The effective physical diffusion coefficient becomes:
\begin{equation}
    D_{u,phys}^* = \frac{D_{u,raw} \times \alpha}{\lambda^2} \approx \frac{(1.52 \times 10^{-10}) \times 1.34}{2.56} \approx 7.956 \times 10^{-11}\text{ m}^2/\text{s}
\end{equation}
Substituting this into the normalization formula yields the dimensionless diffusion coefficient for $A\beta$:
\begin{equation}
    D_{u,norm} = \frac{D_{u,phys}^* \times T}{L^2} \approx \frac{(7.956 \times 10^{-11}) \times 31,557,600}{0.167^2} \approx 0.090028
\end{equation}

In contrast, the experimental diffusion coefficient for Tau ($v$) is reported as $D_{v,raw} = 14.4\text{ }\mu\text{m}^2/\text{s} = 1.44 \times 10^{-11}\text{ m}^2/\text{s}$ \cite{janning2014single}. Because this measurement was obtained at physiological body temperature ($37^\circ\text{C}$), the temperature correction is unnecessary. The effective physical diffusion coefficient with only the tortuosity correction applied is:
\begin{equation}
    D_{v,phys}^* = \frac{D_{v,raw}}{\lambda^2} \approx \frac{1.44 \times 10^{-11}}{2.56} \approx 5.625 \times 10^{-12}\text{ m}^2/\text{s}
\end{equation}
Consequently, the dimensionless diffusion coefficient for Tau is calculated as:
\begin{equation}
    D_{v,norm} = \frac{D_{v,phys}^* \times T}{L^2} \approx \frac{(5.625 \times 10^{-12}) \times 31,557,600}{0.167^2} \approx 0.006365
\end{equation}

\section{Computational Domain and Numerical Methods}

\subsection{Computational Domain and Brain Mesh Representation}

To numerically simulate the spatiotemporal dynamics governed by the Gray-Scott system, we establish a computational domain based on human brain anatomy. Although the human brain is a highly complex three-dimensional volume, a common and effective approach in macroscopic neuro-modeling is to investigate the progression of $A\beta$ and Tau along the cerebral cortex \cite{dan2026uncover,lefevre2010reaction,wei2026neurodetour}. While Alzheimer's pathology actively involves subcortical structures and white matter pathways, the cortical gray matter eventually bears a significant burden of the macroscopic accumulation that correlates with clinical cognitive decline \cite{braak1991neuropathological}. Therefore we restrict our computational domain to a two-dimensional cortical surface manifold extracted from 3D anatomical data.

This surface-bound formulation provides distinct advantages from both biological and computational perspectives:
\begin{itemize}
    \item \textbf{Biological Relevance:} Restricting the domain to the cortical surface reflects the gray matter-centric nature of macroscopic $A\beta$ and Tau accumulation. Furthermore, it facilitates alignment with standard surface-based neuroimaging post-processing pipelines, which reconstruct 3D volumetric scans into 2D surface meshes to analyze cortical thickness and protein density \cite{fischl2012freesurfer,fischl1999cortical,li2012consistent}.
    \item \textbf{Computational Efficiency:} Discretizing a 2D manifold embedded in 3D space reduces the total number of computational nodes compared to a full 3D volumetric mesh. This simplification eliminates the need to define complex internal boundaries for white matter anisotropy, lowering computational cost while maintaining high-resolution representation of the cortical folding \cite{weickenmeier2019physics}.
\end{itemize}

To implement this domain computationally, the continuous cortical surface is discretized into a high-resolution triangular mesh. To accurately solve the governing reaction-diffusion system over this complex geometry, we employ a Finite Element Method (FEM) framework \cite{dhatt2012finite,lee2011stable,xu2017algebraic}. Within this variational formulation, the standard spatial Laplacian operator $\Delta$ is naturally generalized to the Laplace-Beltrami operator $\Delta_{\mathcal{M}}$ to account for diffusion along the curved Riemannian manifold \cite{dziuk2013finite}. By utilizing the discrete triangular elements to approximate the solution space, this FEM approach allows us to rigorously model the geometric influence of the brain's highly folded surface topography on the spatial distribution of the disease end-states.

\begin{figure}[htbp]
    \centering
    \includegraphics[width=0.95\textwidth]{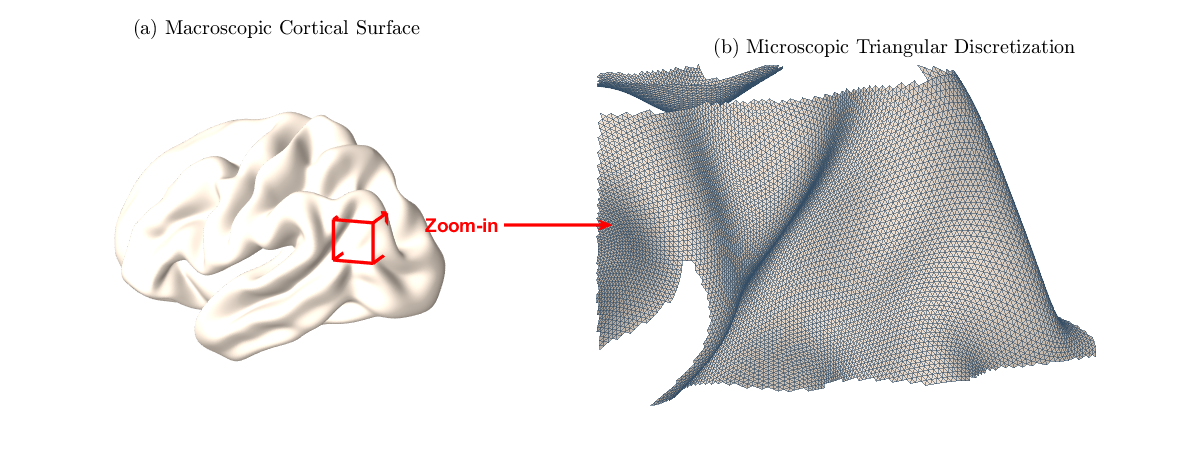} 
    \caption{Computational domain for the Gray-Scott reaction-diffusion system. (a) The macroscopic cortical surface representing the human gray matter manifold, with a bounding box highlighting a representative section of the geometry. (b) A highly resolved microscopic view of the unstructured triangular discretization within this bounded section.}
    \label{fig:brain_domain}
\end{figure}

Figure~\ref{fig:brain_domain} illustrates the geometric configuration of the computational domain alongside its corresponding numerical discretization. This anatomically realistic geometry is constructed using a surface mesh obtained from the Alzheimer's Disease Neuroimaging Initiative (ADNI) dataset \cite{jack2008alzheimer} (\url{https://adni.loni.usc.edu}). Specifically, Figure~\ref{fig:brain_domain}(a) displays the global macroscopic geometry of the human cerebral cortex, capturing the intricate folding patterns of gyri and sulci that characterize the gray matter architecture. To explicitly demonstrate the structural density of the discrete framework, a representative section of the surface is bounded by a red cube and magnified in Figure~\ref{fig:brain_domain}(b). This microscopic view exposes the high-density, unstructured triangular mesh utilized for the spatial discretization. By partitioning the continuous manifold into these fine triangular elements, the local metric tensor of the curved surface is accurately preserved. This underlying mesh fidelity is mathematically essential for the robust evaluation of the Laplace-Beltrami operator $\Delta_{\mathcal{M}}$, ensuring that the simulated spatiotemporal propagation and Turing-type pattern formation of $A\beta$ and Tau are governed by realistic anatomical topography rather than numerical artifacts of an oversimplified domain.

\subsection{Multi-Level Finite Element Framework}
To compute the solutions of the nonlinear governing equations over the discretized cortical manifold $\mathcal{M}$, we develop a robust numerical framework. Given the highly nonlinear nature of the Gray-Scott reaction-diffusion system, the underlying system can exhibit multiple steady-states and complex spatial patterns. Standard iterative solvers often fail to capture this diverse solution space due to severe initial-guess sensitivity. To address this challenge and ensure computational efficiency, we first introduce the standard Finite Element Method (FEM) formulation and subsequently extend it to the Companion-Based Multi-Level Finite Element Method (CBMFEM) \cite{hao2024companion}.

\subsubsection{Finite Element Formulation on Manifolds}
We begin by establishing the weak formulation of the governing Gray-Scott equations on the curved Riemannian manifold $\mathcal{M}$. Let $H^1(\mathcal{M})$ denote the standard Sobolev space on the manifold. From a physical and biological perspective, restricting our solutions to $H^1(\mathcal{M})$ ensures that the modeled protein concentrations are square-integrable (representing finite total mass) and possess square-integrable first derivatives, thereby avoiding non-physical infinite concentration gradients.

Multiplying the governing equations for each biochemical species by a test function $\psi \in H^1(\mathcal{M})$ and applying the integration by parts formula (Green's theorem) on manifolds, the Laplace-Beltrami operator $\Delta_{\mathcal{M}}$ for a concentration $u$ is transformed as follows:
\begin{equation}
    \int_{\mathcal{M}} \Delta_{\mathcal{M}} u \, \psi \, d\mathcal{M} = -\int_{\mathcal{M}} \nabla_{\mathcal{M}} u \cdot \nabla_{\mathcal{M}} \psi \, d\mathcal{M},
\end{equation}
where $\nabla_{\mathcal{M}}$ represents the manifold gradient, and the boundary terms vanish due to the closed, boundary-less nature of the cortical surface manifold \cite{dziuk2006finite}.

Let $V_h \subset H^1(\mathcal{M})$ be a finite-dimensional subspace spanned by piecewise linear basis functions $\{\phi_i\}_{i=1}^N$ associated with the high-resolution triangular elements defined in the computational domain. The continuous concentration of each species is approximated by a discrete expansion $u_h \in V_h$ \cite{ciarlet2002finite}:
\begin{equation}
    u_h(\mathbf{x}, t) = \sum_{i=1}^N U_i(t) \phi_i(\mathbf{x}).
\end{equation}
Substituting these approximations for both interacting species into the variational form spatially discretizes the original partial differential equations (PDEs). This transformation yields a large-scale, strongly coupled discrete dynamical system that governs the transient spatiotemporal evolution of the biochemical network. For steady-state analyses, the temporal derivatives vanish, reducing the formulation to a coupled system of nonlinear algebraic equations, intrinsically expressed in terms of the global mass and stiffness matrices.

\subsubsection{Companion-Based Multi-Level Finite Element Method (CBMFEM)}
Computing multiple solutions for nonlinear differential equations is often challenging because standard iterative solvers, such as Newton's method, are sensitive to initial conditions and typically converge to a single solution. To address this limitation, we employ the Companion-Based Multi-Level Finite Element Method (CBMFEM) to systematically generate robust initial guesses. Rather than replacing standard root-finding algorithms, CBMFEM serves as an efficient initialization framework that enables the global Newton solver to capture multiple distinct solution branches.

The CBMFEM leverages a hierarchical mesh-refinement approach to reduce computational overhead. The algorithm proceeds through the following sequential stages:
\begin{enumerate}
    \item \textbf{Coarsest Grid Initialization:} The solver is initially deployed on the coarsest finite element mesh. Due to the minimal degrees of freedom at this base level, the method focuses solely on capturing the simple, fundamental steady-state solutions of the reaction-diffusion system.
    \item \textbf{Interpolation and Local Companion-Matrix Resolution:} These base solutions are interpolated onto the next hierarchically refined mesh. Instead of solving the entire nonlinear system globally, the companion-matrix-based eigensolver is strategically applied only to the newly introduced spatial nodes. By exploiting polynomial approximations of the local nonlinearities, this step systematically computes the distinct local solution branches at these specific new points.
    \item \textbf{Global Fine-Grid Correction:} The companion matrix constructs multiple localized solutions depending on the degree of the polynomial. Considering the combinations of these local solutions yields a large number of high-quality initial guesses across the entire fine grid. Starting from this extensive pool of initial guesses, a global Newton-type solver is applied to correct and refine them into valid global solutions satisfying standard convergence criteria.
\end{enumerate}

These refined configurations then serve as the new base solutions for the next hierarchical level. While the algorithm can repeat this cyclic process indefinitely, applying the local companion-matrix resolution across increasingly dense meshes incurs significant computational cost. The core advantage of the CBMFEM lies in its ability to systematically capture diverse nonlinear solution branches at coarser and intermediate grid levels. Once a sufficient multiplicity of distinct steady states has been established, the necessity to generate further multiple initial guesses via the companion matrix diminishes. To accelerate the computation on the finer grids, the algorithm adaptively transitions to standard interpolation. The established solutions are directly interpolated onto the higher-resolution mesh and subsequently refined using the global Newton solver. By leveraging the companion matrix primarily where it is most impactful---to uncover the fundamental solution landscape early in the hierarchy---this approach significantly reduces computational time while maintaining the quality and diversity of the final steady states, avoiding the prohibitive expense of solving the global problem on highly dense grids.

Furthermore, while the current numerical implementation is explicitly tailored to the Gray-Scott model, the underlying CBMFEM framework possesses broader mathematical applicability. The methodology can be directly extended to a wider class of two-component reaction-diffusion equations that exhibit Turing-type instabilities. The primary algebraic requirement for this extension is that the reaction kinetic functions, $F(u,v)$ and $G(u,v)$, are polynomial in nature, and that their nonlinear terms can be decoupled or canceled out through a linear combination. Under these specific conditions, the local companion-matrix formulation can systematically resolve the distinct solution branches independent of the specific kinetic system. Moreover, while our present numerical execution relies on a Newton solver to converge these branches, the framework's modularity allows it to be readily integrated with various alternative nonlinear solvers \cite{abdelrahman2020robust,lang2013adaptive,lee2025parallel,lee2006new,lo2012robust}. Consequently, this hierarchical approach provides a generalized computational strategy for discovering multiple coexisting steady states across various polynomial-based reaction-diffusion systems commonly encountered in mathematical biology, such as the Schnakenberg, Brusselator, and FitzHugh-Nagumo models.

\subsubsection{Construction of the Multi-Level Mesh Hierarchy}
The implementation of the CBMFEM inherently requires a nested sequence of triangular meshes, ranging from a coarse base level to the highly resolved target manifold. However, standard neuroimaging pipelines typically output a single, high-density surface mesh representing the cortical geometry. Since a predefined coarse grid is not naturally given, we systematically construct the required multi-level hierarchy through a top-down mesh coarsening strategy.

Starting from the original high-resolution cortical mesh, coarser grids are generated by selectively removing a subset of vertices. Following this vertex removal, the remaining nodes are locally retriangulated to form a valid, continuous surface. This point-removal process is carefully controlled to progressively reduce the degrees of freedom while preserving the essential topological features and the macroscopic curvature of the brain surface.

\begin{figure}[htbp]
    \centering
    \includegraphics[width=1.0\textwidth]{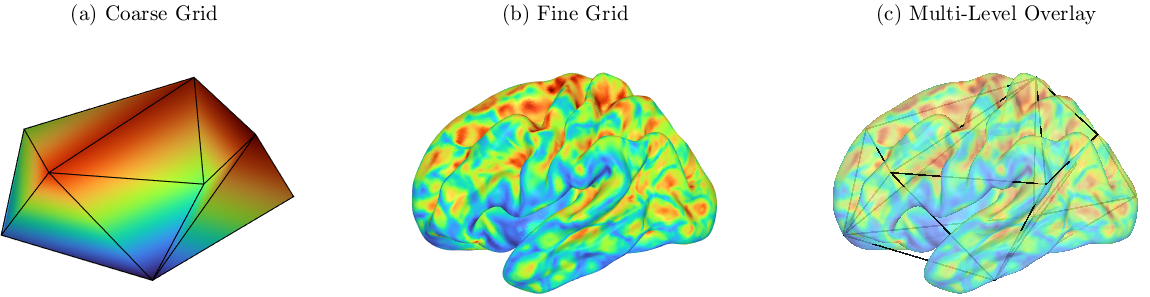} 
    \caption{Multi-level mesh hierarchy constructed for the CBMFEM framework. For illustrative clarity, only two representative levels are compared here from the full multi-step refinement sequence. (a) A representative coarse grid generated via vertex decimation. (b) The target high-resolution fine grid representing the continuous cortical manifold. (c) An overlay of the coarse edges on the fine surface, demonstrating the strictly nested structural alignment between consecutive refinement levels.}
    \label{fig:mesh_hierarchy}
\end{figure}

Figure~\ref{fig:mesh_hierarchy} visually demonstrates this hierarchical relationship and the resulting structural framework. Figure~\ref{fig:mesh_hierarchy}(a) displays a representative coarse grid generated through our decimation process, which serves as the computational foundation for identifying the base steady-state solutions with minimal computational overhead. Conversely, Figure~\ref{fig:mesh_hierarchy}(b) illustrates the high-resolution fine grid that accurately captures the complex anatomical folding of the target manifold. The spatial alignment between these discrete levels is explicitly shown in the multi-level overlay in Figure~\ref{fig:mesh_hierarchy}(c). Because the decimation strategy ensures that the vertices of the coarse grid are a strict subset of the fine grid, the meshes are perfectly nested. This exact spatial overlap provides the necessary structural foundation for the hierarchical interpolation and guarantees that the solver can unambiguously isolate the newly added nodes to execute the local companion-matrix resolution steps of the CBMFEM.

\section{Simulation Results}
Before applying the proposed numerical framework to the complex topography of the 3D cortical manifold, we first present numerical experiments on a simplified two-dimensional planar domain. For this 2D investigation, we adopt a pre-computed brain mesh originally developed and provided in a previous study \cite{ju2020parameter}, which was generated utilizing GMSH software. Rather than modifying the geometry, our primary objective is to use this established planar domain to rigorously benchmark and validate the performance of our Companion-Based Multi-Level Finite Element Method (CBMFEM) in capturing the highly nonlinear dynamics of the Gray-Scott reaction-diffusion system.

\subsection{2D Domain}
In this preliminary 2D test case, we apply isotropic and spatially constant diffusion coefficients, adopting previously calculated values for both interacting proteins. While this serves as a basic simplification—given that diffusivity in the actual human brain is highly heterogeneous and anisotropic, differing substantially between the dense white matter tracts and the cortical gray matter regions—it is appropriate for the strict purpose of algorithmic verification. Utilizing these constant coefficients allows us to isolate the performance of the nonlinear solver from the complexities of spatially varying parameters.

\begin{figure}[htbp]
    \centering
    \includegraphics[width=1.0\textwidth]{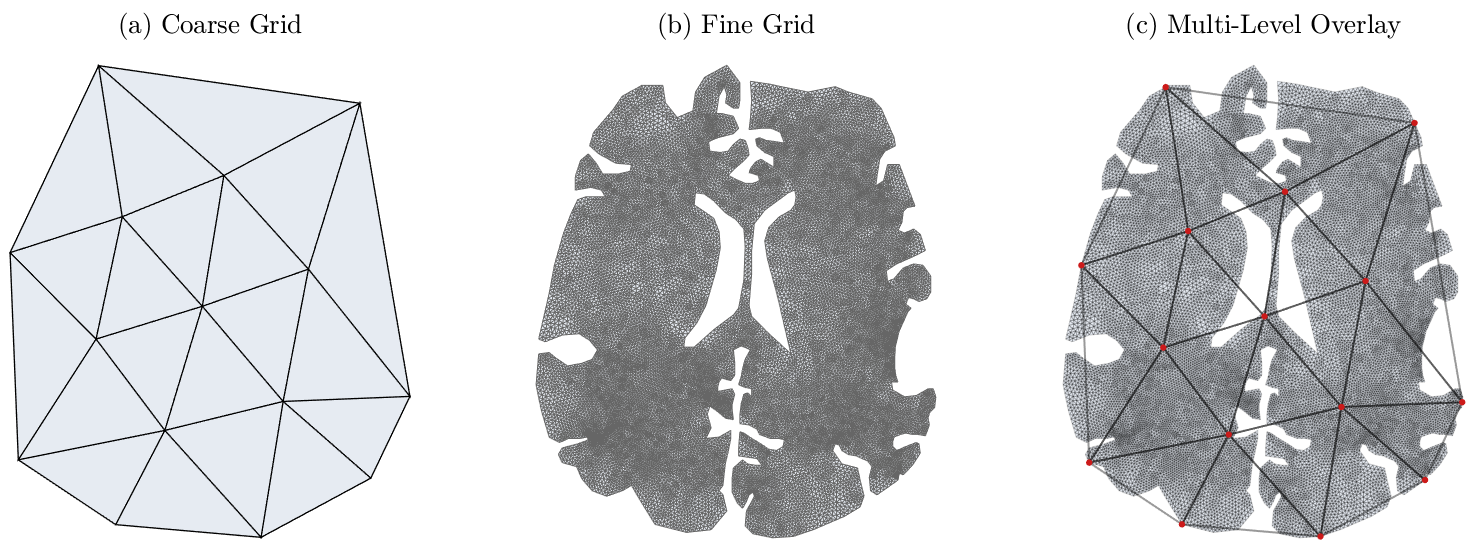}
    \caption{Two-dimensional multi-level mesh hierarchy for algorithmic validation. (a) The foundational coarse grid. (b) The target high-resolution fine grid. (c) An overlay demonstrating the strictly nested structural alignment, with coarse nodes (red) highlighted to indicate the initialization points for the local companion-matrix solver.}
    \label{fig:2d_mesh}
\end{figure}

To execute the CBMFEM algorithm on this planar domain, we construct a nested multi-level sequence of 2D meshes. Figure~\ref{fig:2d_mesh} illustrates the structural hierarchy employed for this validation. Figure~\ref{fig:2d_mesh}(a) shows a representative coarse grid, which provides an appropriately reduced number of degrees of freedom to efficiently compute the base steady-state solutions. Figure~\ref{fig:2d_mesh}(b) depicts the high-resolution fine grid required to capture the intricate details of the reaction-diffusion patterns. As demonstrated by the multi-level overlay in Figure~\ref{fig:2d_mesh}(c), the coarse and fine grids are strictly nested.

\begin{figure}[htbp]
    \centering
    \includegraphics[width=1.0\textwidth]{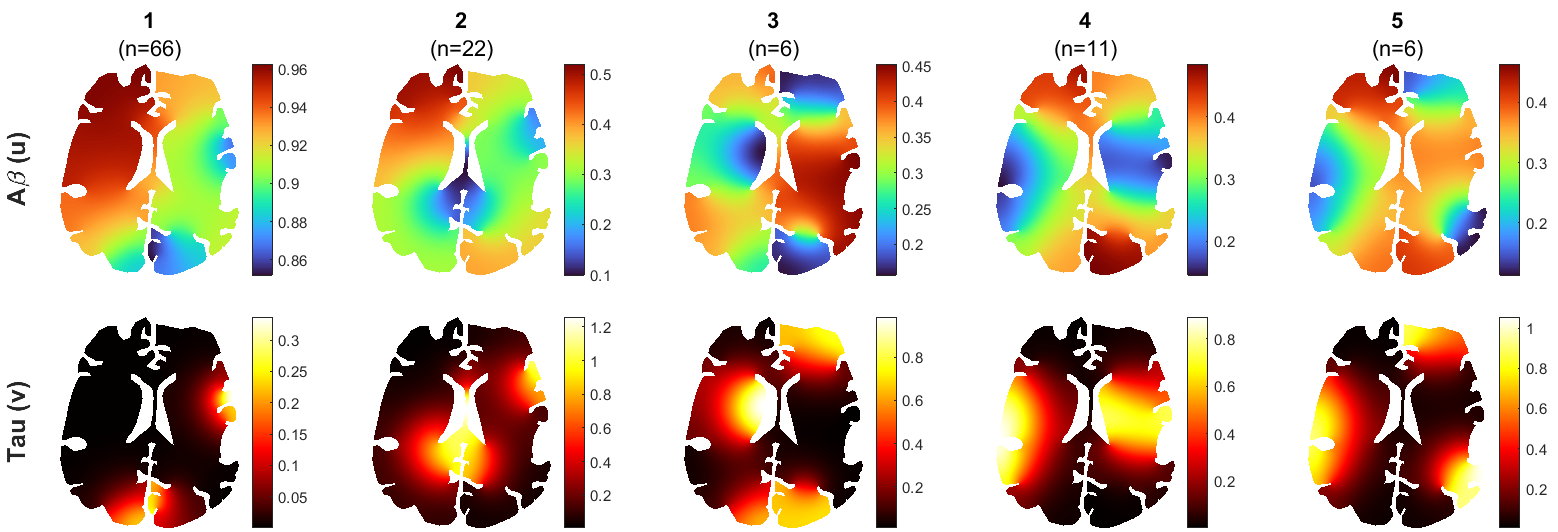}
    \caption{Representative steady-state patterns of the Gray-Scott model on a 2D domain. The set of 111 numerical solutions was optimally partitioned into 5 distinct fundamental phenotypes using PCA and Silhouette-evaluated $K$-means clustering.}
    \label{fig:all_clusters}
\end{figure}

A key phenomenon of interest in this benchmark is the emergence of complex spatial configurations, specifically the diverse pattern formation characteristic of the Gray-Scott dynamics. Through the execution of the CBMFEM framework over this hierarchical 2D domain, we successfully isolated a total of 111 distinct steady-state solutions. Rather than analyzing each spatial configuration as an isolated entity, we systematically categorized these diverse pattern formations using Principal Component Analysis (PCA) coupled with $K$-means clustering. To ensure the statistical reliability and robustness of this classification, we employed Silhouette analysis to determine the optimal number of phenotypic groups. As illustrated in Figure~\ref{fig:all_clusters}, this rigorous evaluation mathematically partitioned the solution space into 5 fundamental representative pattern types.

\begin{figure}[htbp]
    \centering
    \includegraphics[width=0.95\textwidth]{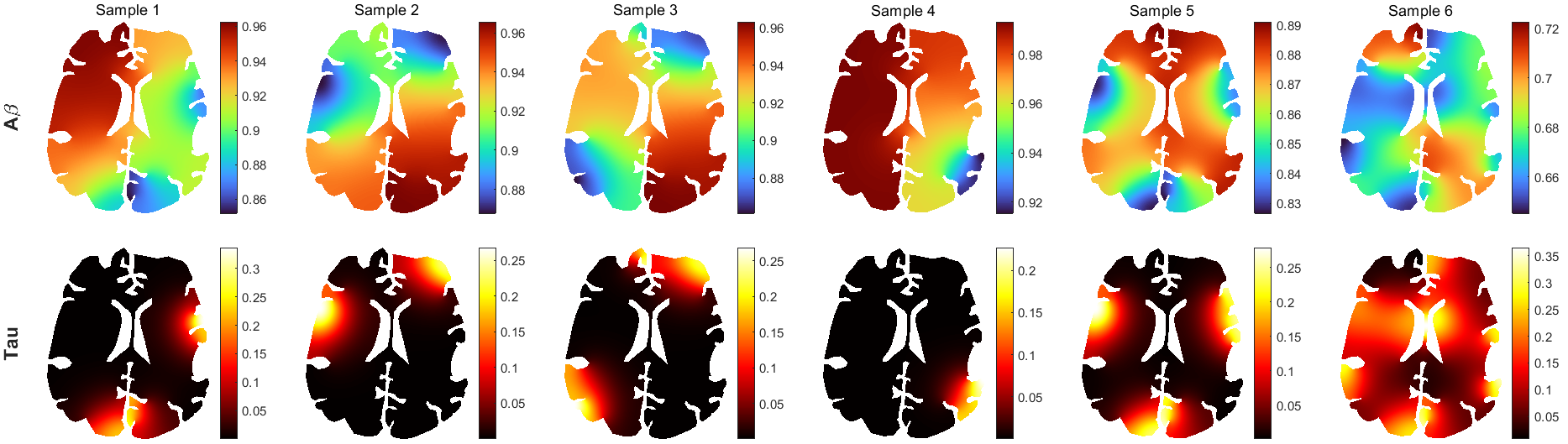}
    \caption{Validation of intra-cluster morphological consistency. The figure displays six randomly selected steady-state solutions from Fig. \ref{fig:all_clusters} Pattern Type 1, visually demonstrating the high degree of structural similarity and pattern fidelity within the identified cluster.}
    \label{fig:type1_samples}
\end{figure}

To further validate the morphological consistency of our data-driven classification, Figure~\ref{fig:type1_samples} visually demonstrates the intra-cluster similarity by displaying six randomly sampled spatial configurations exclusively from Pattern Type 1. The structural uniformity within this group confirms that our clustering methodology successfully groups solutions based on their fundamental macroscopic architectures, effectively looking past minor spatial shifts or superficial variations. Overall, the ability to robustly generate, systematically categorize, and accurately resolve these complex Turing-type patterns on a 2D domain verifies the reliability of the CBMFEM framework.

Crucially, the success of this PCA and $K$-means clustering pipeline extends beyond mere 2D algorithmic verification. It demonstrates that highly complex reaction-diffusion dynamics can be systematically distilled into a finite set of distinct structural phenotypes. Consequently, this validation not only establishes the necessary computational confidence to project our simulations onto the geometrically intricate, fully curved 3D cortical surface, but also highlights the broader clinical potential of our methodology. By applying this exact data-driven classification to 3D anatomical manifolds, our framework lays a robust theoretical foundation for identifying macroscopic pathological patterns in the brain, a critical conceptual step toward the future pattern-based classification and potential diagnosis of neurodegenerative diseases.

Building upon this theoretical foundation, the specific topographic profiles of each cluster type offer biological and clinical insights, translating mathematical spatial mismatch into predictable patient symptomatology. Type 1, the most frequent phenotype ($n=66$), closely aligns with the classic preclinical or asymptomatic stage of Alzheimer's disease \cite{sperling2011toward}. In this configuration, $A\beta$ exhibits a heavy, widespread neocortical burden (consistently high at 0.86--0.96), yet the downstream Tau pathology remains remarkably low (confined below 0.3). According to the amyloid cascade hypothesis, amyloid accumulation occurs decades upstream of actual neurodegeneration \cite{jack2010hypothetical}. Therefore, despite the extensive global amyloid plaque burden, the relative absence of aggressive Tau tangles suggests that cognitive functions, memory, and executive control may remain largely intact in these individuals, mathematically representing a clinically silent trajectory of the disease.

In contrast, Type 2 ($n=22$) exhibits marked hemispheric asymmetry---A$\beta$ heavily saturates the left hemisphere, whereas Tau preferentially accumulates in the right. Because Tau pathology drives actual neurodegeneration and correlates much more closely with clinical deficits than A$\beta$ \cite{ossenkoppele2016tau}, a hypothetical patient with this profile would likely suffer from right-hemisphere-dominated symptoms, such as visuospatial disorientation. Conversely, left-hemisphere functions, such as language, could remain relatively preserved despite the high A$\beta$ burden, highlighting the critical paradigm that local Tau, rather than A$\beta$, dictates regional neurotoxicity \cite{galton2000atypical, graff2021new}.

Types 3, 4, and 5 represent additional structural variants characterized by localized Tau hotspots and spatially discordant A$\beta$ distributions. Type 3 ($n=6$) exhibits a different lateralization pattern compared to Type 2. In this configuration, A$\beta$ is more prominent in the right hemisphere, while elevated Tau is primarily localized to the left. Clinically, such left-hemisphere-predominant Tau pathology can be associated with language-variant presentations, such as primary progressive aphasia (PPA), where patients often experience language deficits while retaining relatively intact visuospatial skills. 

Types 4 ($n=11$) and 5 ($n=6$) display a distinct spatial configuration with a more symmetric lateralization. In these variants, A$\beta$ accumulation is more prominent at the anterior and posterior poles, whereas Tau pathology is largely distributed along the lateral cortical margins, structurally corresponding to the temporal and parietal regions. Notably, the frontal and occipital areas in these types remain relatively spared from Tau despite a high A$\beta$ presence. Biologically, this topographic distribution---where Tau is primarily found in the lateral temporoparietal networks while relatively sparing the poles---resembles the typical progression pathways seen in the intermediate stages of Alzheimer's disease \cite{braak1991neuropathological}. These varied spatial configurations demonstrate that complex anatomical geometry can support diverse spatial patterns, providing a potential theoretical link to the heterogeneity observed in clinical phenotypes.

\subsection{3D Cortical Surface}

Building upon the successful algorithmic verification in the 2D domain, we extend the application of the CBMFEM framework to the geometrically complex, fully curved 3D manifold of the human cerebral cortex, using a surface mesh obtained from the Alzheimer's Disease Neuroimaging Initiative (ADNI) dataset \cite{jack2008alzheimer} (\url{https://adni.loni.usc.edu}). Unlike the planar test case, the actual cortical surface is characterized by highly irregular topography, including intricate folding patterns of gyri and sulci. Simulating reaction-diffusion dynamics on such a heterogeneous domain is notoriously challenging, yet it is essential for accurately modeling the macroscopic propagation and accumulation of pathological proteins in the human brain.

Applying the CBMFEM algorithm to the high-resolution 3D cortical mesh, we generated a comprehensive set of steady-state solutions. To eliminate redundant numerical convergence artifacts and ensure the integrity of our dataset, we applied a strict $L^\infty$-norm filtering criterion. This preprocessing step successfully isolated a refined set of 36 strictly unique spatial distributions. 

\begin{figure}[htbp]
    \centering
    \includegraphics[width=1.0\textwidth]{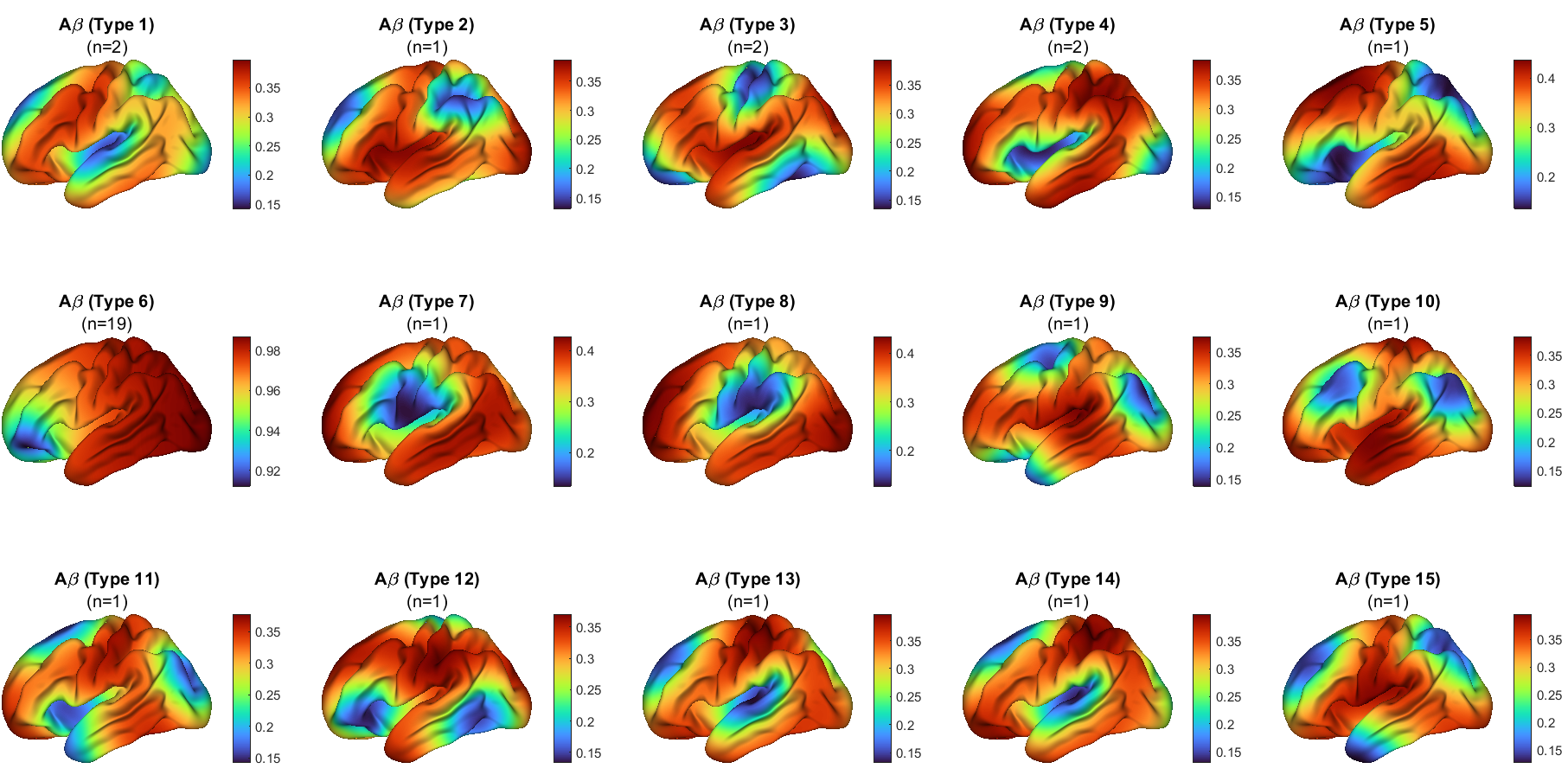}
    \includegraphics[width=1.0\textwidth]{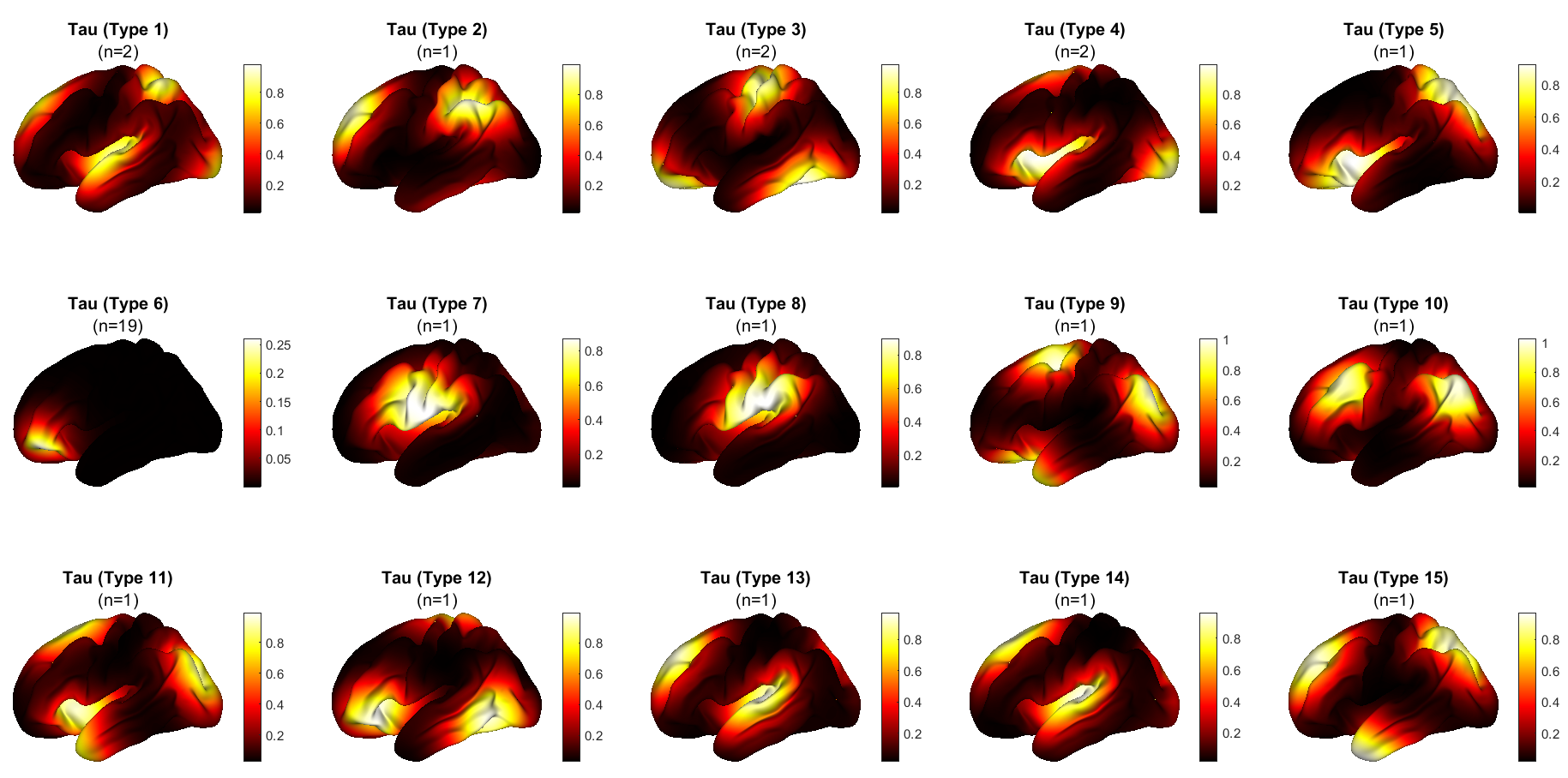}
    \caption{Representative 3D steady-state pattern formations on the human cortical surface. The figure displays 15 distinct structural phenotypes clustered from 36 unique solutions, illustrating the complex reaction-diffusion dynamics across the intricate topography of gyri and sulci.}
    \label{fig:3d_main_clusters}
\end{figure}

Consistent with our methodology established in the 2D validation, we subsequently analyzed this high-dimensional solution space using PCA coupled with $K$-means clustering. Evaluated via Silhouette analysis to maximize mathematical variance ratios, the algorithm optimally partitioned the 36 unique solutions into 15 distinct phenotypic clusters. As illustrated in Figure~\ref{fig:3d_main_clusters}, this data-driven pipeline successfully categorized the fundamental macroscopic configurations directly on the 3D anatomical manifold.

Consistent with our methodology established in the 2D validation, we subsequently analyzed these 36 unique solutions using PCA coupled with $K$-means clustering. Evaluated via Silhouette analysis, the algorithm partitioned the solution space into 15 distinct phenotypic clusters. As illustrated in Figure~\ref{fig:3d_main_clusters}, this data-driven pipeline successfully categorized the fundamental macroscopic configurations directly on the 3D anatomical manifold. Given the high diversity of the generated solutions---with several clusters representing highly unique, localized pattern formations---we present set of 15 representative centroids to capture the spectrum of spatial variations without redundant intra-cluster validation.

Ultimately, the successful generation and categorization of these diverse pattern formations on the 3D cerebral cortex represent a significant computational milestone. By demonstrating that highly nonlinear biochemical dynamics can be reliably simulated and distilled into distinct structural phenotypes on realistic human anatomy, this framework provides a powerful theoretical tool for exploring the diverse pathological end-states of neurodegenerative diseases.

The transition from the 2D cross-section to the fully folded 3D cortical manifold intrinsically introduces geometric complexity. Interestingly, while our 2D simulations generated a larger pool of raw numerical solutions (111 solutions) that converged into merely 5 macroscopic clusters, the 3D simulations distilled a considerably smaller set of raw solutions (36 solutions) into a significantly higher number of distinct structural phenotypes (15 clusters). It is important to acknowledge that the absolute total number of existing stable solutions in this highly nonlinear system remains unknown. Although we utilized the CBMFEM to systematically discover multiple steady states, an exhaustive search was precluded by current computational limits. Thus, the smaller pool of raw 3D solutions might simply be a consequence of the substantially higher computational expense of fully 3D curved surface simulations, though it remains an open mathematical question whether the 3D manifold inherently harbors fewer total stable states.

Regardless of the raw solution count, the expansion in macroscopic diversification---from 5 clusters in the 2D domain to 15 in 3D---suggests a significant role of geometric complexity. The intricate topology of the sulci and gyri introduces highly variable boundary and non-uniform distances across the cortical manifold. Consequently, rather than the numerical solutions being statistically grouped into a few broad macroscopic patterns during clustering, the dynamics on the 3D surface are partitioned into a wider array of localized, distinct equilibria. Biologically, this mathematically derived diversification, potentially influenced by the underlying physical geometry, may provide a conceptual parallel to the vast clinical and neuroanatomical heterogeneity of Alzheimer's disease observed in real-world patient populations.

Within this expanded 3D spectrum, the model identifies a prominent baseline: Type 6 ($n=19$). Notably, over half of the 3D solutions are grouped into this single cluster. Consistent with our 2D findings, this major phenotype portrays a widespread A$\beta$ burden (with concentrations peaking near 0.98) coupled with a relative absence of cortical Tau. Biologically, this phenotype aligns well with the preclinical stage of Alzheimer's disease \cite{jack2010hypothetical, sperling2011toward}. This suggests that a globally amyloid-positive yet tau-negative brain state \cite{jack2018nia} is not merely a geometric artifact, but a stable mathematical equilibrium that can persist even on complex anatomical geometries.

Conversely, the remaining variant clusters exhibit focal, regionally specific Tau aggregations. Rather than spreading uniformly across the cortex, Tau pathology in these 3D variants appears to localize into discrete anatomical regions---such as the temporal pole, specific frontal gyri, or parietal lobules. This localized distribution of Tau provides a mathematical parallel to distinct neurodegenerative trajectories~\cite{vogel2021four}. In vivo, Tau pathology is known to propagate through specific cortical networks, leading to localized cognitive deficits depending on the anatomical epicenter~\cite{ossenkoppele2016tau}. By generating these heterogeneous Tau hotspots out of a limited pool of raw solutions, our 3D framework suggests a potential geometric context for this phenomenon: the inherent curvature and physical topology of the human brain may act as a structural scaffold, potentially influencing biochemical dynamics to form diverse, clinically relevant disease variants~\cite{graff2021new}.

Crucially, the highly localized nature of these Tau hotspots naturally suggests the existence of extensive cortical areas that remain largely spared from severe Tau pathology, despite a pervasive A$\beta$ burden. Biologically, these mathematically stable regions with minimal Tau serve as a mathematical analogue for the ``insensitive'' or resistant regions of the human brain. Clinical observations confirm that specific functional areas---such as the primary sensory, motor, and visual cortices---exhibit strong resistance to Tau propagation and remain functionally intact until the terminal stages of the disease \cite{ aksman2023data,braak1991neuropathological}. 

By confining severe Tau accumulation to restricted anatomical regions, these specific reaction-diffusion patterns represent a mathematically stable state that mirrors limited disease progression. Because the pathology is geographically contained rather than globally distributed, these configurations may correspond to the biologically plausible states that allow some individuals to maintain preserved cognitive function despite substantial AD pathology. This spatial containment provides a theoretical basis for the therapeutic paradigm discussed subsequently: rather than solely aiming for the complete eradication of pathological proteins, interventions could strategically steer the disease dynamics toward these localized, relatively benign stable states.

\section{Conclusion}

In this work, we developed a reaction-diffusion framework to investigate the interactions between Amyloid-$\beta$ ($A\beta$) and Tau proteins and their potential role in pattern formation during Alzheimer's disease progression. Motivated by the clinical observation that some individuals exhibit substantial levels of $A\beta$ and Tau pathology while remaining cognitively normal, we hypothesized that the spatial organization of pathological proteins may be as important as their overall burden. To explore this hypothesis, we formulated a Gray-Scott type reaction-diffusion model, implemented it on both two-dimensional domains and anatomically realistic cortical surface meshes, and employed the Companion-Based Multi-Level Finite Element Method (CBMFEM) to systematically compute multiple steady-state solutions.

Our numerical results revealed a rich landscape of stable spatial patterns supported by the coupled $A\beta$--Tau system. Rather than converging to a single pathological endpoint, the model admits numerous distinct steady-state configurations. Through dimensionality reduction and clustering analyses, these solutions were organized into a finite collection of representative pattern phenotypes, demonstrating that the underlying disease dynamics possess multiple potential long-term outcomes. The existence of these stable states suggests that Alzheimer's disease progression may not be governed solely by the amount of pathological protein accumulation, but also by the spatial structure and stability of protein interactions across the brain.

From a biological perspective, these findings provide a potential mathematical explanation for the well-documented heterogeneity observed in Alzheimer's disease. In particular, the presence of multiple stable patterns raises the possibility that certain configurations of $A\beta$ and Tau may be relatively benign, allowing individuals to tolerate substantial pathological burden without developing severe cognitive impairment. Conversely, other configurations may promote widespread propagation of pathology and accelerate neurodegeneration. While the present study does not establish direct clinical correspondence between individual patterns and patient outcomes, it provides a theoretical framework for investigating such relationships in future studies.

The proposed framework also suggests an alternative perspective on therapeutic intervention. Current disease-modifying therapies primarily aim to eliminate pathological proteins from the brain. In contrast, our results indicate that treatment may instead be viewed as a control problem in which disease dynamics are guided from aggressive pathological states toward more favorable and stable configurations. Under this paradigm, the ultimate therapeutic objective is not necessarily the complete removal of $A\beta$ and Tau, but rather the stabilization of disease trajectories within regions of the solution landscape associated with reduced neurotoxicity and preserved cognitive function.

Several important directions remain for future research. The present model considers a simplified interaction mechanism between $A\beta$ and Tau and assumes homogeneous diffusion on the cortical surface. Future work will incorporate additional biological processes, including neuronal loss, neuroinflammation, glymphatic clearance, and network-based transport through structural connectivity. Furthermore, integrating patient-specific neuroimaging data from large-scale cohorts such as ADNI may enable the identification of clinically relevant pattern phenotypes and facilitate personalized prediction of disease progression. Finally, combining the pattern formation framework with optimal control and reinforcement learning approaches may provide a principled methodology for designing individualized treatment strategies that actively steer disease dynamics toward favorable stable states.

Overall, this study demonstrates that reaction-diffusion theory and pattern formation analysis provide a promising mathematical framework for understanding the complex spatial dynamics of Alzheimer's disease. By revealing the existence of multiple stable pathological configurations, our work offers a new perspective on disease heterogeneity, cognitive resilience, and personalized therapeutic intervention. We believe that characterizing the landscape of stable $A\beta$--Tau patterns represents an important step toward a more mechanistic and predictive understanding of Alzheimer's disease progression.

\section*{Acknowledgment}

S.L. and W.H. are supported by both NIH via 1R35GM146894 and NSF DMS-2533995.  W. H. was also supported by the Huck Chair in AI Mathematical Modeling from Penn State University's Huck Institutes of the Life Sciences.

\section*{Statements and Declarations}

\textbf{Competing Interests}\\
The authors declare that they have no conflict of interest.\\
\textbf{Data Availability}\\
Data sharing is not applicable to this article as no new experimental datasets were generated or analyzed during the current study.

\bibliographystyle{plain}
\bibliography{references}

\end{document}